\documentclass{PoS}

\title{Results from the first missions of the JEM-EUSO program}

\ShortTitle{Results from the first missions of the JEM-EUSO program}

\author{\speaker{Francesco Fenu} for the JEM-EUSO Collaboration\footnote{for collaboration list see PoS(ICRC2019)1177} \\ \\
        Universit\`a degli studi di Torino, INFN Torino\\
        E-mail: \email{francesco.fenu@gmail.com}}

\abstract{The origin of Ultra-High Energy Cosmic Rays (UHECRs) remains unsolved in contemporary astroparticle physics.
  The objective of the JEM-EUSO program is the realization of a space mission devoted to UHECR physics.
  Several detectors have been developed in this framework or are under development.
  EUSO--TA, installed at the Telescope Array site in Utah in 2013, is in operation.
  It has detected 9 UHECRs in coincidence with Telescope Array fluorescence detector at Black Rock Mesa.
  EUSO--Balloon flew on board a stratospheric balloon in August 2014.
  It measured the UV intensity on forests, lakes and cities as well as proved the observation of UHECR--like events reproduced by laser tracks.
  EUSO--SPB1 was launched on board a super pressure balloon on April 24th 2017 and flew for 12 days.
  It proved the functionality of all the subsystems of the telescope on a long term and observed the UV emission on oceans.
  TUS, on board the Lomonosov satellite, in orbit since April 28th 2016, is now included in the JEM--EUSO program and has detected so far in the UHECR trigger-mode a few interesting signals.
  Mini--EUSO is in its final phase of integration in Russia, and  is scheduled for launch on August 22$^{nd}$.
  The main results obtained so far by these missions are summarized and put in the perspective of future space detectors such as K--EUSO and POEMMA.}

\FullConference{36th International Cosmic Ray Conference -ICRC2019-\\
		July 24th - August 1st, 2019\\
		Madison, WI, U.S.A.}

\begin{document}

\section{Introduction}
The origin of Ultra High Energy Cosmic Rays (UHECR) remains unsolved in contemporary astrophysics.
The very low fluxes at such extreme energies, of the order of few particles per $\mathrm km^2$, sr, century are a challenge for current observatories.
The main goal of the Joint Experiment Mission--EUSO (JEM--EUSO) program is the investigation of the UHECRs of the most extreme energy through the detection of the fluorescence light emitted by
Extensive Air Showers in the atmosphere \cite{JEMEUSO_Program}.
From several hundreds of kilometers of altitude, with a wide Field Of View (FOV), such telescopes will allow an extremely
large exposure and therefore will probe the most extreme part of the spectrum.

The program consists of several missions, some of them under development, others in operation and others already done in the past.
EUSO--TA is currently taking data since 2015 at the Black Rock Mesa site of the Telescope Array observatory \cite{EUSO-TA}.
Two stratospheric balloon flights have been performed in the past: the EUSO--Balloon \cite{EUSO-Balloon} for one night in 2014 and the EUSO--SPB1 \cite{EUSO-SPB} for 12 days in 2017. 
A follow--up mission, EUSO--SPB2 \cite{EUSO-SPBII}, is in preparation for a third balloon flight in 2021--22.
The Mini-EUSO detector \cite{mini-EUSO} is planned to fly on August 22$^{nd}$ 2019 onboard the International Space Station (ISS) to map the emission in the UV from space.
The TUS telescope onboard the Lomonosov satellite \cite{TUS} is in orbit since 2016 and has taken  data for 45 days from the altitude of 550 km.
K-EUSO \cite{K-EUSO} is planned to fly in 2023 and to detect UHECR showers from space from the Russian segment of the ISS.
The POEMMA observatory \cite{POEMMA} will instead be a free flier telescope planned to take data from 2029.

In this contribution we will show the main results obtained in the framework of this program and put them in context of the future missions, K--EUSO and POEMMA.

\section{EUSO--TA}
\begin{figure}[t]
\def\figh{0.33}
\centering
\includegraphics[height=5.9cm]{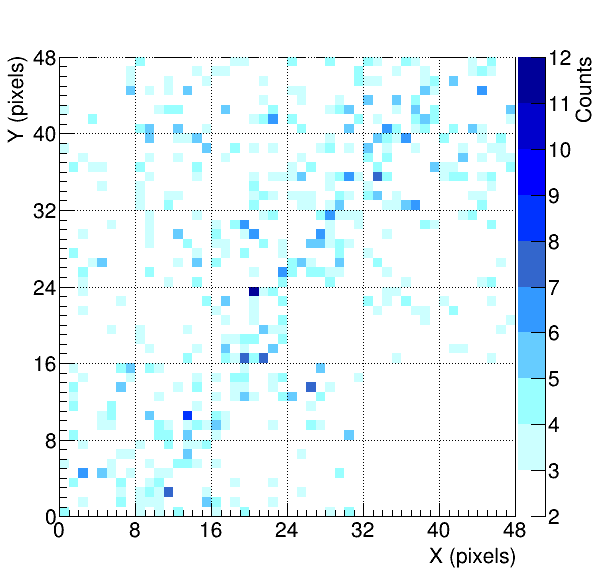}\hfill
\includegraphics[height=5.5cm]{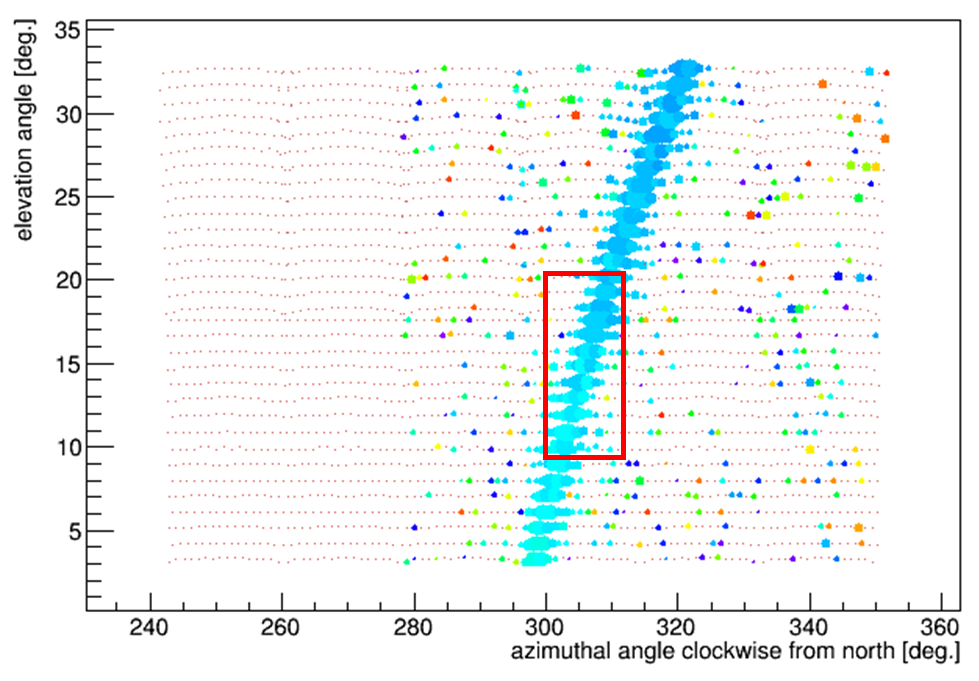}
\caption{the image of a $\sim 10^{18}$ eV event, with 2.5km impact parameter, detected in coincidence by the EUSO--TA (left) and the TA--BRM telescope (right).}
\label{fig:exampleImage}
\end{figure}

The EUSO--TA telescope is in operation since 2015 in front of the Black Rock Mesa (BRM) site of the TA observatory \cite{piotrowsky}\cite{Bisconti19}.
The detector uses the external trigger provided by the BRM telescope to detect cosmic ray showers.
A total of 120 hours of data have been used in the present analysis and nine EAS events have been detected.
An example of such events can be seen in Fig. \ref{fig:exampleImage} where the EUSO--TA image can be seen on the left while the same event as observed by
BRM is shown on the right. The small FOV of the EUSO-TA telescopes allows the detection of only a partial fraction of the shower and the independent reconstruction of the
shower parameters is therefore not possible.
\begin{figure}[h!]
\centering
\includegraphics[width=0.7\textwidth]{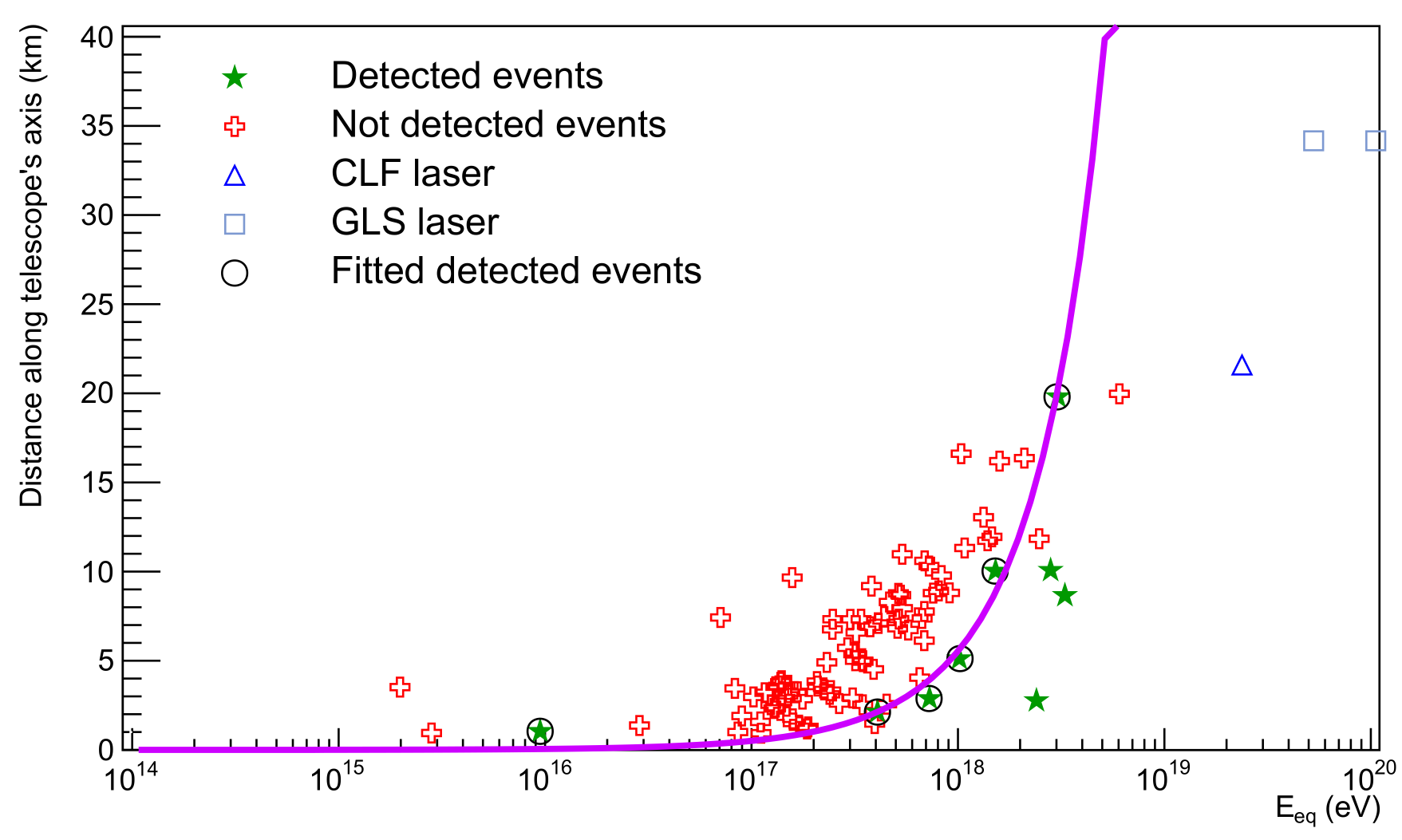}
\caption{the EUSO--TA detection maximal distance as a function of energy.}
\label{fig:Efficiency-EUSOTA}
\end{figure}
The shower parameters reconstructed by the TA collaboration have been therefore used in this analysis.

To estimate the exposure of the future missions it is fundamental to estimate the energy threshold of the EUSO--TA telescope as a function of distance.
For this purpose we therefore use the events detected and reconstructed by TA.
In Fig. \ref{fig:Efficiency-EUSOTA} are shown the events observed by the BRM telescope in the vicinity of the EUSO-TA FOV.
All the showers have been reconstructed and a correction to the event energy has been applied to take into account the fact that EUSO--TA does not necessarily point to the shower maximum.
Such corrected energy has been calculated through Montecarlo using the shower parameters from the TA reconstruction and the EUSO--TA orientation.
This parameter, defined as equivalent energy ($\mathrm E_{eq}$), has been put on the X axis.
On the Y axis we observe the distance of the shower along the telescope longitudinal axis.
As red crosses we see the events that were not visible despite being in the EUSO--TA FOV, while as green stars we depict the events that have been identified inside the EUSO--TA data.
As can be seen, the only events which are visible are the ones on the bottom right side of the plot (i.e. close and energetic showers) while most of the showers did not trigger.
The fit on the lowest energy triggering events with a second degree polynomial is shown as a continuous line. Such a function gives a parametrical expression for the
energy threshold that can be rescaled for other detectors of the JEM-EUSO program. We also see in the plot the very bright CLF and GLS laser events. Such events are on the far right of
the plot since their luminosity is enough for an unambiguous detection.

Another approach to calibrate the detector is shown in \cite{plebaniak}. Stars are used here as reference sources to calibrate the detector.
EUSO--TA is going to be upgraded in the near future with a new electronics and an autonomous trigger system \cite{battisti} and a digital
camera is currently being operated on the site to study meteors and probe the speculative models of nuclearites \cite{ide}.

\section{EUSO--Balloon}
The EUSO--Balloon detector was launched by CNES on August $25^{th}$ 2014 from the Timmins base in Ontario (Canada).
Such detector flew for one night at the floating altitude of 38 km over different kinds of landscapes.
The telescope was not equipped with a trigger electronics and was therefore acquiring data at a fixed rate.
A total amount of over 2.5 millions images of the FOV have been therefore registered.
A wide variety of background situations has been encountered: cities and other man--made sources, forests, lakes and clouds.
The detector mapped the UV emission from the ground in the 290--430 nm band \cite{kenjiEUSO-Balloon}.
\begin{figure}[h!]
\centering
\includegraphics[width=0.7\textwidth]{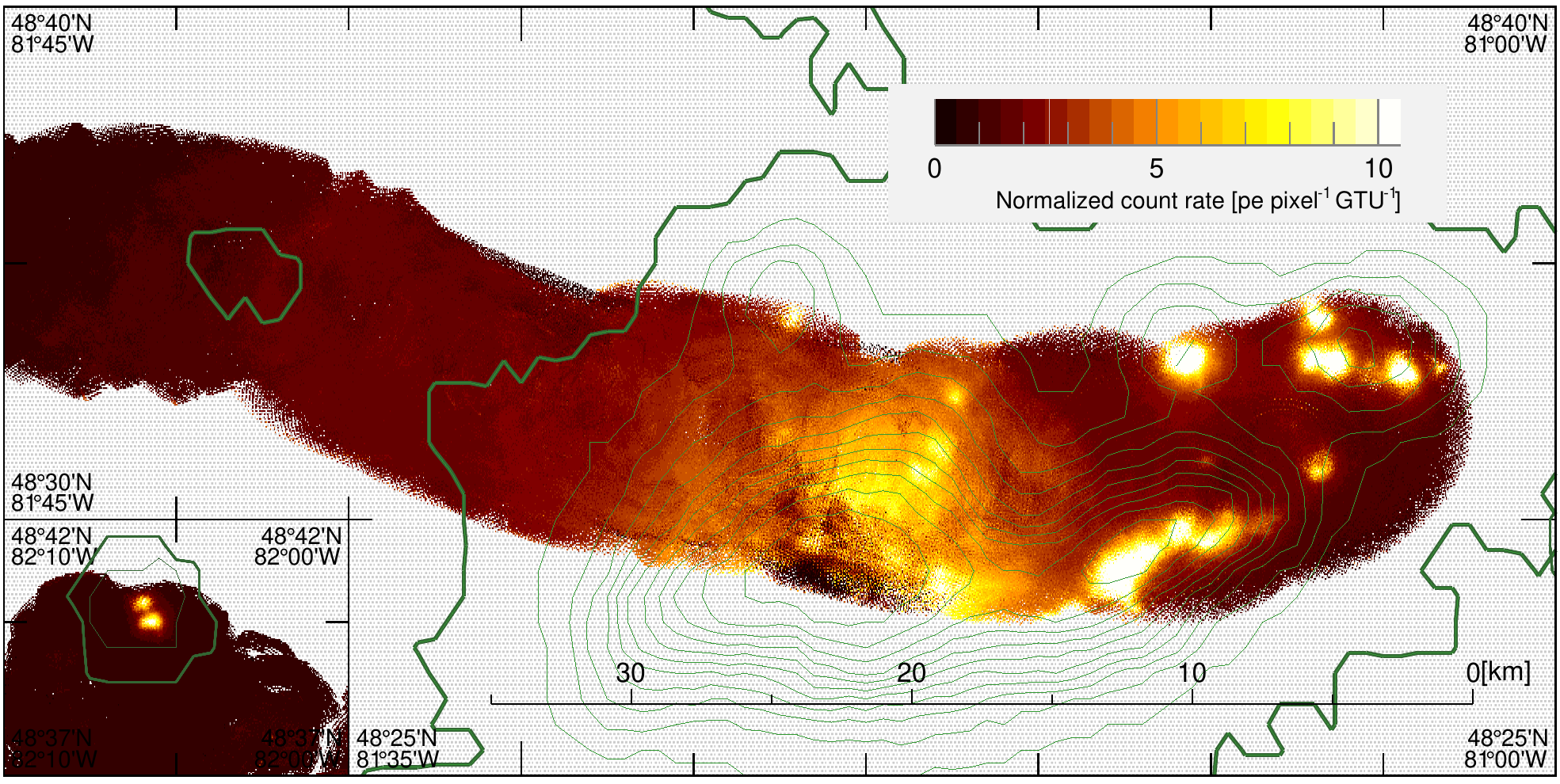}
\caption{the EUSO--Balloon map of background. }
\label{fig:EUSO-BALLOON}
\end{figure}
The average count rate per pixel in the clear sky condition, without artificial sources has been estimated to be of the order of $\sim$ 0.65 counts / pixel / GTU\footnote{Gate Time Unit, the frame of the JEM--EUSO detector (2.5 $\mathrm \mu s$) }. In presence of clouds a background a factor $\sim$ 2 higher has to be expected while this rate increases by an order of magnitude over urban areas.
Such measurements, to be regarded as valid only for balloon flights, are important in the exposure estimation of the future space based detectors.
Under specific assumptions of starlight and airglow spectrum, and thanks to detector simulations,  a photon radiance of the order of 300--320 photons / ns / $\mathrm m^2$
has been estimated for the clear sky condition and without artificial light. These values are however strongly affected by the assumptions on the
airglow spectrum, detector efficiency and can be regarded as pure indications on the radiance values that has to be expected.

\begin{figure}[t]
\def\figh{0.33}
\centering
\includegraphics[height=4.3cm]{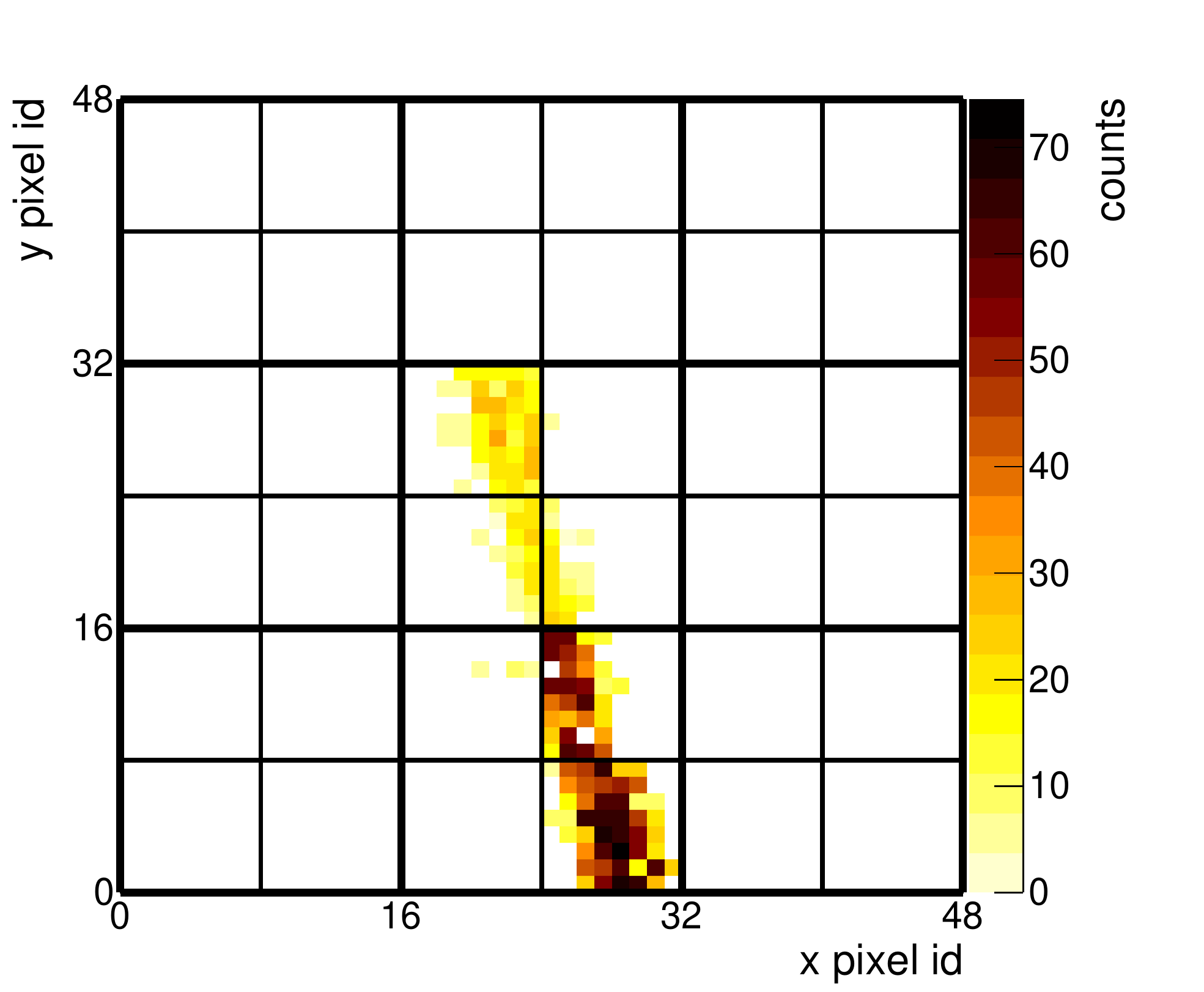}\hfill
\includegraphics[height=4cm]{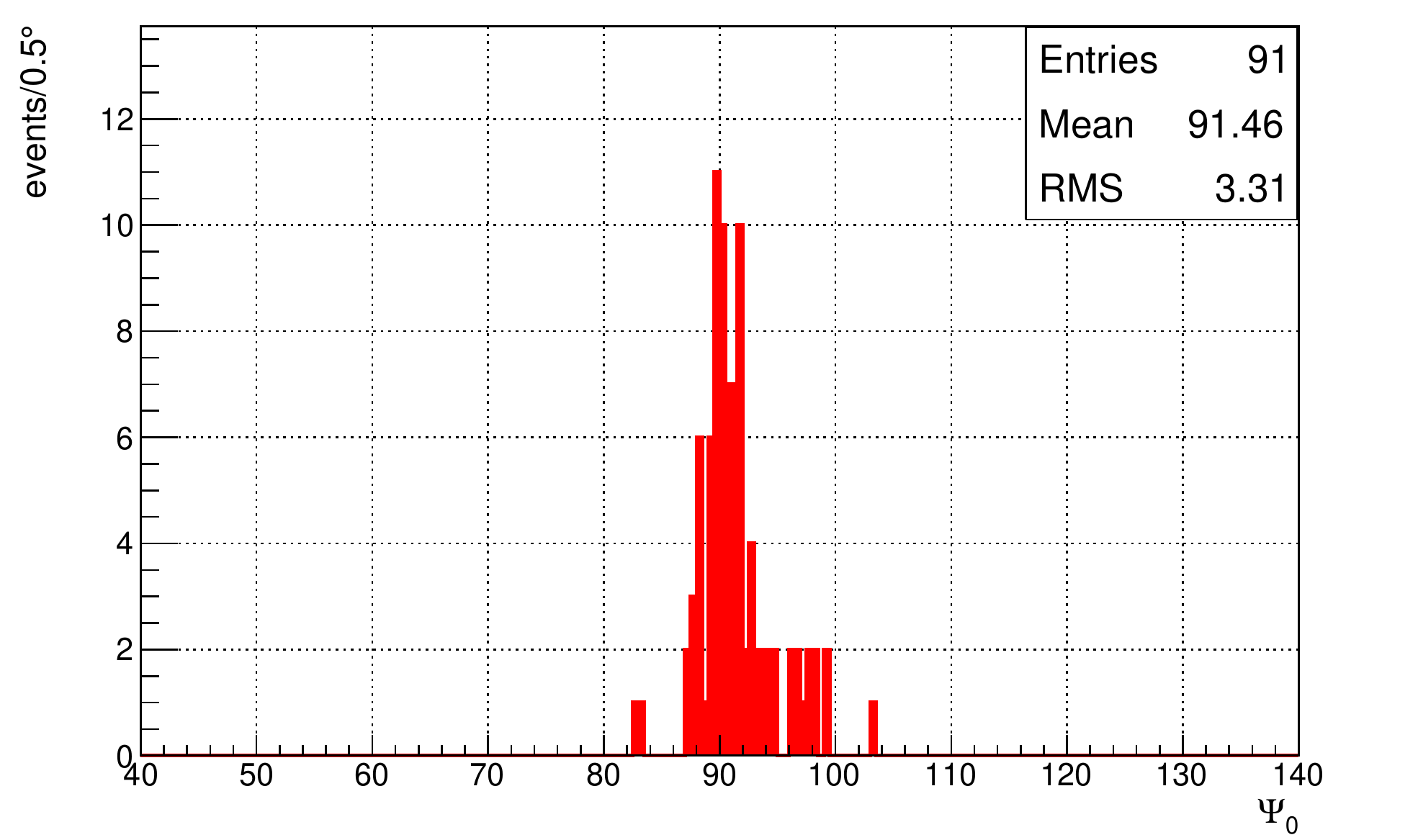}
\caption{the reconstruction of the direction of the laser pulse as obtained from the EUSO--Balloon data.}
\label{fig:angularRecoBalloon}
\end{figure}

At the time of flight an helicopter circled along the trajectory of the balloon at an altitude of 3 km. A laser, which emitted pulses of light horizontally, was installed on board the helicopter.
Thanks to scattering in the atmosphere, the laser pulses could be imaged by EUSO--Balloon as bright spots moving on the focal surface at the speed of light, mimicking therefore a cosmic ray
shower signal.
Several hundreds laser events have been successfully acquired. Such events were used to test the reconstruction capabilities of the detector.
The laser pulse direction was reconstructed thanks to a time fit applied on the data \cite{Eser18}. The energy reconstruction was however not possible given the
very high energy of the pulse and the strong saturation of the detector.

\section{EUSO--SPB1}
EUSO--SPB1 was launched on April $25^{th}$ 2017 from Wanaka, New Zealand as a mission of opportunity on a NASA Super Pressure Balloon (SPB) test flight.
The flight was planned to circle the southern hemisphere, following high altitude winds, for a duration of the order of 100 days and was designed
to detect cosmic rays. The detector was therefore equipped with an autonomous trigger \cite{bayer17}, a set of hard discs where to store the events and an antenna for the transmission of the data.
A SiPMT camera was installed on the payload to test the readiness of such detector on stratospheric balloon conditions \cite{painter19}.

\begin{figure}[h!]
\centering
\includegraphics[height=4.3cm]{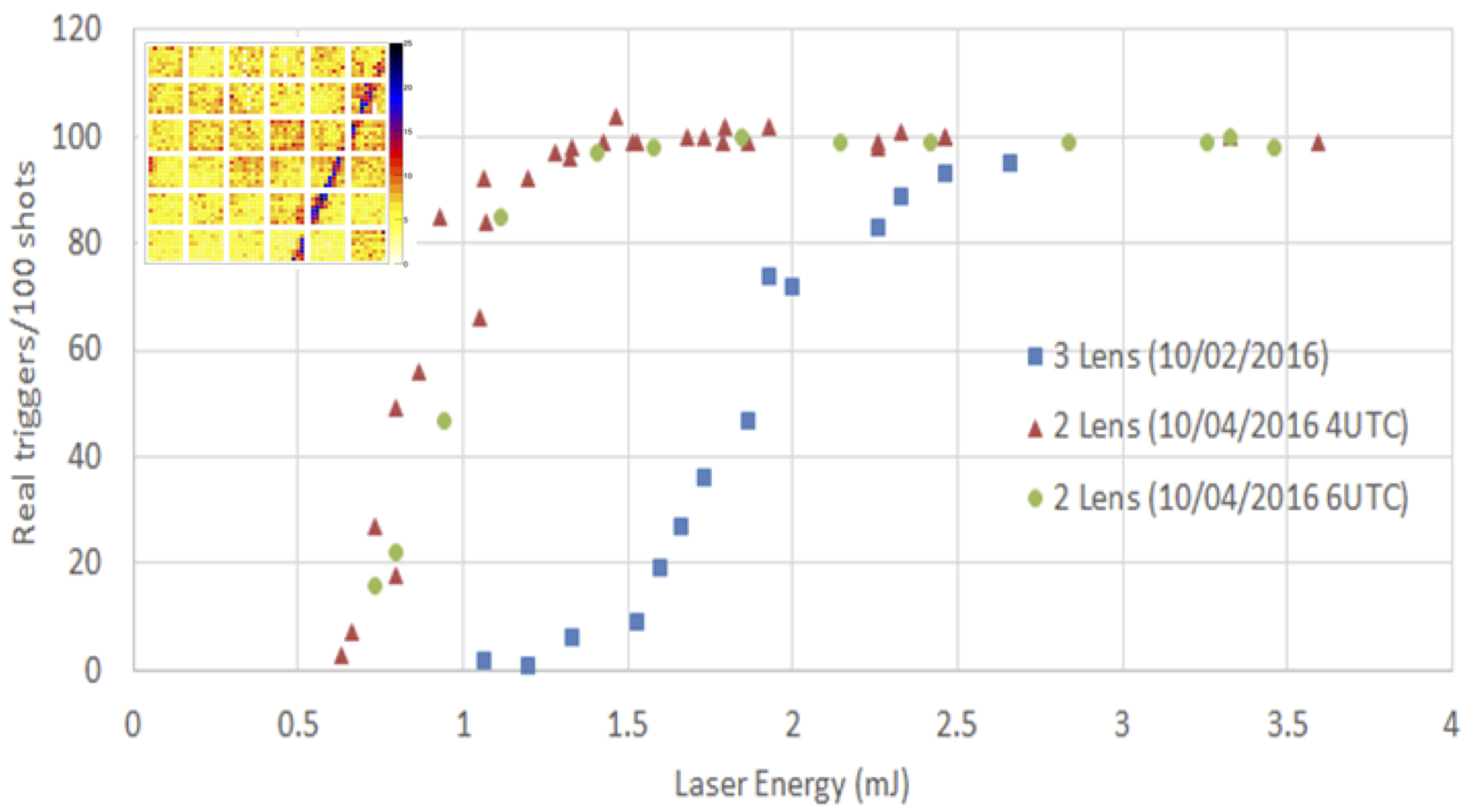}\hfill
\includegraphics[height=4.3cm]{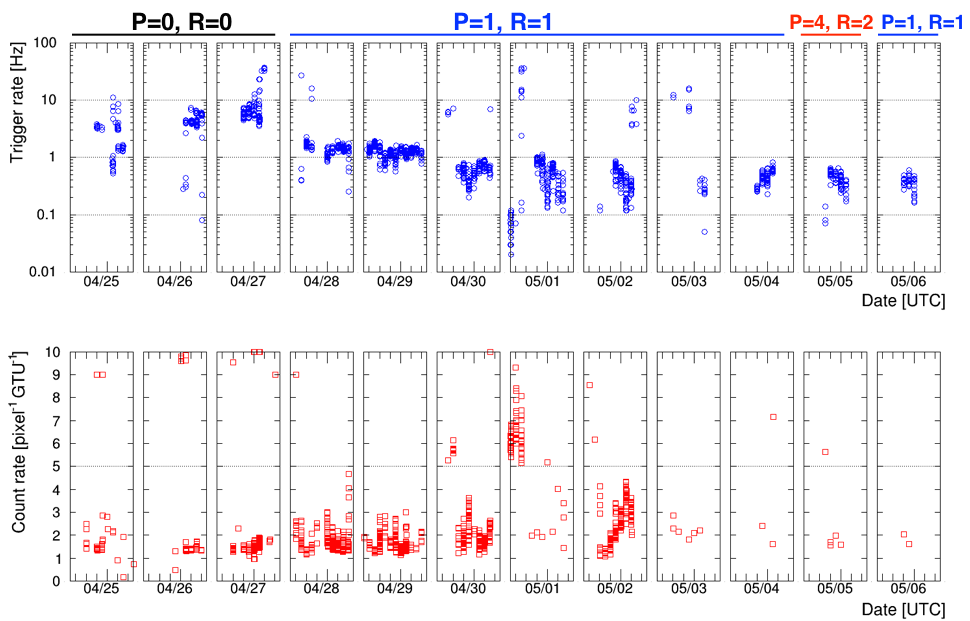}

\caption{on the left the efficiency curve of the EUSO--SPB detector as function of laser energy.
  On the right the trigger rate (upper panel) and the background rate per pixel per GTU (lower panel).}
\label{fig:Efficiency-EUSO-SPB}
\end{figure}

The telescope has been tested on the TA site in Utah in October 2016, to test its functionality  and to estimate the trigger energy  threshold.
The detector operated for several nights and detected laser events, stars and meteors. The 50$\%$ efficiency has been estimated to be at
$\sim 3 \cdot 10^{18}$ eV for a balloon at 33 km altitude for a 45 degrees shower through the analysis of several laser pulses.
Several detector configurations have been tested, in order to maximize the
detector throughput. In Fig. \ref{fig:Efficiency-EUSO-SPB} the optical configuration with two or three lenses can be seen. The testing at the TA site allowed to
prove that the two lenses configuration has a better efficiency compared to the three lens system.

After 12 days of flight the balloon developed a leak and sunk into the ocean thus strongly reducing the available exposure. The hard discs, have been lost and the entire analysis had
to be performed with the data transferred with the antenna.
Up to now, no cosmic ray candidate has been therefore found in the EUSO--SPB data.
Studies on the background rate \cite{battistiNIM} were fundamental to determine the exposure curve as a function of time and to give an estimate of the number of expected events.
Simulation studies performed after the flight proved that an average of 1 event is to be expected over the entire flight (which consisted of $\sim$ 30 hours of scientific data) \cite{shinozakiSPB}.
The search of cosmic ray events is still ongoing through both traditional \cite{diaz19} and machine learning approaches\cite{vrabel}.

\section{TUS e Mini--EUSO}
\begin{figure}[h!]
\centering
\includegraphics[width=0.4\textwidth]{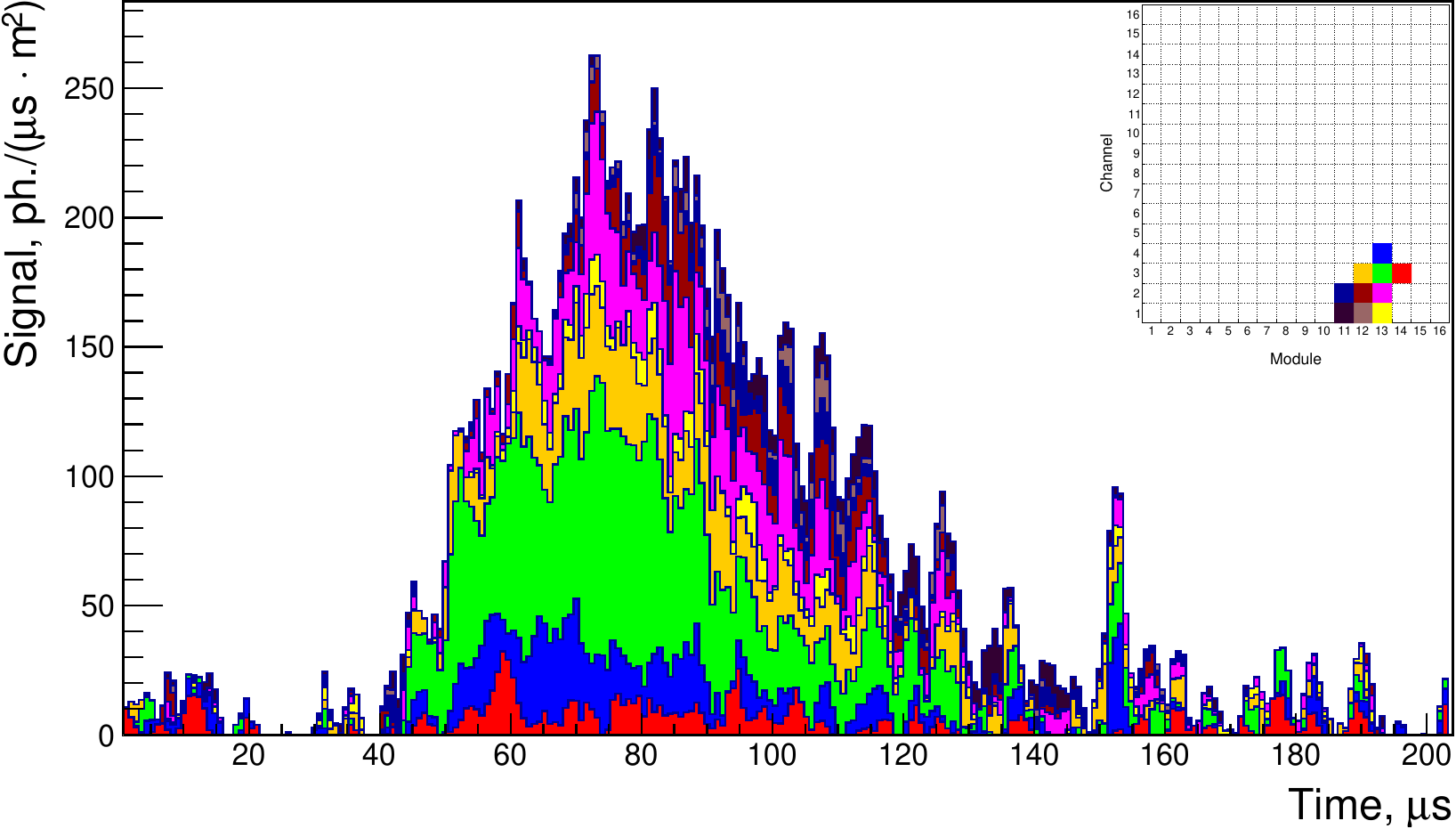}
\caption{one of the events detected by TUS and that are currently under analysis.}
\label{fig:TUS-example}
\end{figure}

The Track Ultraviolet Setup has been launched on April 28 2016 onboard the Lomonosov satellite.
TUS is the first space based detector aiming at the detection of cosmic ray showers from space through
fluorescence.
The instrument was taking data until November 2017 and different acquisition modes were tested: cosmic ray, lightning and meteor mode.
Preliminary analysis of the cosmic ray mode data show the presence of several 10$^{-5}$ s scale events which resemble some of the features of a cosmic ray shower.
Such events (one example is presented in Fig. \ref{fig:TUS-example}) are still under study and a more detailed analysis is ongoing \cite{Zotov2019}.

TUS detected also meteors, planes, cities, lightnings and other atmospheric transient events.
The study of such events is of fundamental importance to estimate possible sources of background for future space experiments.
Moreover, data acquired in meteor mode are used to constrain physics beyond the standard model like nuclearites \cite{shinozakiNuclearites}.

The Mini--EUSO detector will fly on August 22$^{nd}$ 2019 on the Russian segment of the ISS and will monitor the atmosphere from a UV transparent window.
From 400 km the Mini--EUSO detector will measure the background from the same altitude as the future space based detectors.
Balloons cannot detect the direct airglow emission which is instead detectable from space. It is therefore necessary to fly a prototype at
orbital altitude to be able to give a good estimate of the exposure of space based detectors.
\begin{figure}[h!]
\centering
\includegraphics[width=1\textwidth]{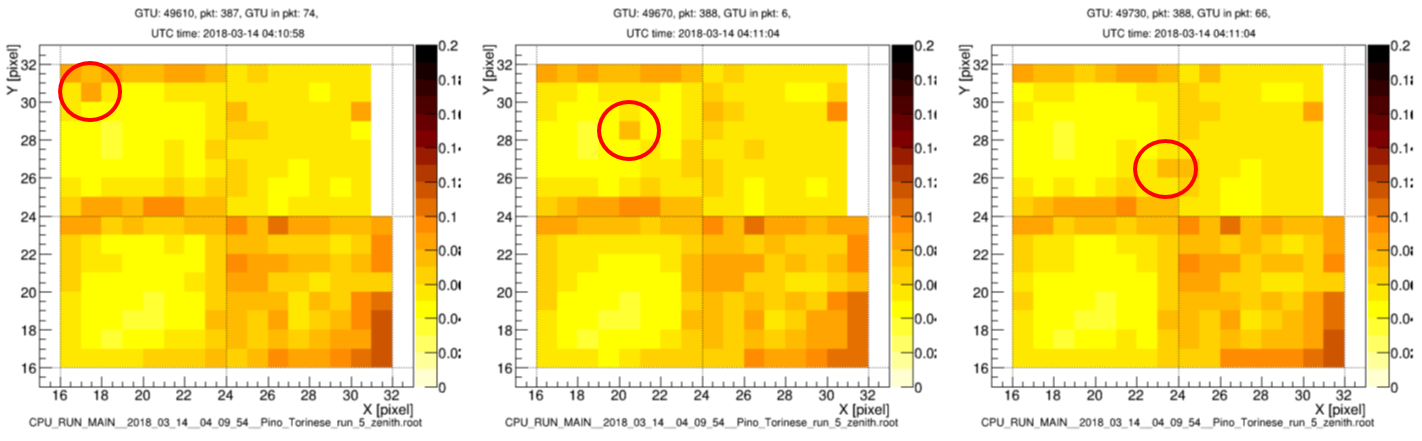}
\caption{the track of the Meteor 1--31 rocket body detected by the Mini--EUSO detector from Torino on March 14 2018.}
\label{fig:Mini-EUSOsatellite}
\end{figure}
The Mini--EUSO detector will be also tested through ground laser sources \cite{kungel} and will also detect many slow events like lightnings, meteors and debris.
The Mini--EUSO detector was tested in Torino at the beginning of 2018 and could detect a wide variety of background sources.
In Fig. \ref{fig:Mini-EUSOsatellite} an example of a satellite detected from ground is shown \cite{miyamoto2017}. This allowed to understand the
threshold on the  debris detection and to give an estimate of the expected rate of such events in view of the very close launch date.

 \section*{Acknowledgments}
 This work was partially supported by Basic Science Interdisciplinary Research
Projects of RIKEN and JSPS KAKENHI Grant (JP17H02905, JP16H02426 and
JP16H16737), by the Italian Ministry of Foreign Affairs and International
Cooperation, by the Italian Space Agency through the ASI INFN agreement
n. 2017-8-H.0 and contract n. 2016-1-U.0, by NASA award 11-APRA-0058 in
the USA, by the
Deutsches Zentrum f\"ur Luft- und Raumfahrt, by the French space agency
CNES, the Helmholtz Alliance for Astroparticle Physics funded by the
Initiative and Networking Fund of the Helmholtz Association (Germany), by
Slovak Academy of Sciences MVTS JEMEUSO as well as VEGA grant agency project
2/0132/17,
by National Science Centre in Poland grant (2015/19/N/ST9/03708 and
2017/27/B/ST9/02162), by Mexican funding agencies PAPIIT-UNAM, CONACyT and
the
Mexican Space Agency (AEM). Russia is supported by ROSCOSMOS and the Russian
Foundation for
Basic Research Grant No 16-29-13065. Sweden is funded by the Olle Engkvist
Byggm\"astare Foundation.

\end{document}